\documentclass[aps,pra,twocolumn,superscriptaddress,showpacs,amsmath,amssymb]{revtex4-1}
\usepackage{amsmath}
\usepackage{amssymb}
\usepackage{graphicx}
\usepackage{here}
\usepackage{bm}
\usepackage{color}
\usepackage{braket}

\usepackage{hyperref}

\begin{document}
\title{Magnetic and Electronic Properties of Spin-Orbit Coupled Dirac Electrons \\
on a $(001)$ Thin Film of Double Perovskite Sr$_2$FeMoO$_6$}
\author{Masahiko G. Yamada}
\email{myamada@mp.es.osaka-u.ac.jp}
\affiliation{Department of Materials Engineering Science, Osaka University, Toyonaka 560-8531, Japan}
\affiliation{Institute for Solid State Physics, University of Tokyo, Kashiwa 277-8581, Japan}
\affiliation{Max Planck Institute for Solid State Research, Heisenbergstrasse 1, D-70569 Stuttgart, Germany}
\author{George Jackeli}
\altaffiliation[]{Also at Andronikashvili Institute of Physics, 0177 Tbilisi, Georgia}
\affiliation{Max Planck Institute for Solid State Research, Heisenbergstrasse 1, D-70569 Stuttgart, Germany}
\affiliation{Institute for Functional Matter and Quantum Technologies, University of Stuttgart, Pfaffenwaldring 57, D-70569 Stuttgart, Germany}
\date{\today}

\begin{abstract}
We present an interacting model for the electronic and magnetic behavior of a strained $(001)$ atomic layer of Sr$_2$FeMoO$_6,$
which shows room-temperature
ferrimagnetism and magnetoresistance with potential spintronics application in the bulk.
We find that the strong spin-orbit coupling in the molybdenum 4$d$ shell gives
rise to a robust ferrimagnetic state with an emergent spin-polarized electronic structure consisting of flat bands and four massive or massless Dirac dispersions.
Based on the spin-wave theory, we demonstrate that the magnetic order remains intact for a wide range of doping,
leading to the possibility of exploring flat band physics, such as Wigner crystallization in electron-doped Sr$_{2-x}$La$_{x}$FeMoO$_6.$ \\ \\
PhySH: Monolayer films, Ferrimagnetism, Half-metals
\end{abstract}

\maketitle

\section{Introduction}
The coexistence of a strong spin-orbit coupling (SOC)
and low dimensionality gives rise to novel quantum phases of matter~\cite{Bosh2017}.
Two-dimensional (2D) systems confined in the atomically thin films can possess
rich electronic properties different from the bulk and could host various new
correlated phenomena. Especially, the rapid progress in synthesizing atomic-scale slabs, superlattices and heterostructures of
correlated transition metal oxides by pulsed laser deposition or molecular beam epitaxy has motivated the exploration of various (perovskite) compounds
epitaxially grown on different cubic substrates as potential nano-scale devices~\cite{Hwan2012,Bosh2017,Taka2010,JACE:JACE02556}.
An advantage of the epitaxial
growth is that by changing the substrate, one can introduce strain to thin films due to a mismatch of the lattice constants and
thereby control the electronic state, which we call strain engineering~\cite{Chene1600245}.
By replacing 3$d$ transition metal ions with heavier 4$d$ or 5$d$ ions, one can even
control the strength of the SOC.
These flexibilities of epitaxially grown atomic-scale layers could pave a way to search for unusual spin-orbit coupled correlated phenomena in 2D systems.
In this context, theoretical exploration of possible phases can provide a useful guidance.

In order to explore such collective phenomena,
perovskite oxide is one of the best-established platforms~\cite{Bosh2017}.
For instance, LaAlO$_3$/SrTiO$_3$ (LAO/STO) interface is known as a good 2D electron gas
system~\cite{Ohtomo2004} and hosts various electronic phenomena, including 2D superconductivity~\cite{Reyren1196} and Rashba SOC effects~\cite{Shalom2010}.
Although there have appeared new types of atomic-scale 2D systems like
the transition metal dichalcogenides~\cite{Wang2012}, where we can expect a stronger effect of
SOC than in graphene~\cite{KaneMele}, perovskite systems are still
important playgrounds because the knowledge on their synthesis and chemical properties
has been accumulated for a long time.

Specifically, the bulk double perovskite compounds, such as Sr$_2$FeMoO$_6$ (SFMO) and $A_2\mathrm{FeMoO_6}$ ($A$ = Ca, Ba, Pb), have been investigated intensely as examples of half-metallic
ferrimagnets (FiM) with enhanced magnetoresistance at room temperature
and as possible spintronics
devices based on the high spin polarization of charge
carriers~\cite{SFMOKobayashi,MAIGNAN1999224,AFMO,PFMO}.
Theoretical studies of the carrier induced FiM in cubic SFMO have been previously discussed within the
\textit{ab initio}~\cite{Sarma2000,Fang2001,Solo2002,Sait2002} and model Hamiltonian approaches~\cite{Chat2001,Jackeli,Onur2011}
without including the SOC.

Since double perovskite compounds have two-sublattice structure, the synthesis
of high-quality SFMO thin films with completely staggered Fe/Mo sublattices is experimentally
challenging.
Here, motivated by a recent fabrication of well-ordered
thin films of double perovskite SFMO epitaxially
grown along the $(001)$ direction on various perovskite substrates
that showed a ferrimagnetic ground state~\cite{substrate}, we theoretically explore
the combined effects of the strong SOC, tetragonal elongation, and carrier
doping in a $(001)$ layer of SFMO.
For example, a typical perovskite substrate
STO has a slightly shorter lattice constant ($\sim -1.1$\%)
than SFMO, so STO/SFMO heterostructures~\cite{Hauser} would be ideal systems to investigate such effects.

In the insulating compounds such as $\mathrm{Ba_2YMoO_6}$, the SOC locally stabilizes the $j=3/2$ quartet of Mo$^{5+}$ and triggers rich multiorbital physics~\cite{Balents,Natori2016,Romh2016},
unlike in the insulating iridates, where the orbital shape of the
lowest energy $j=1/2$ state is fixed, and the SOC manifests itself in the
anisotropic exchange interactions~\cite{Chen2008,Jackeli2009}.
On the other hand, in SFMO the molybdenum 4$d$ electrons are itinerant, forming a conduction band. In this case, a strong impact of the SOC
on the band structure is expected. Indeed, it has recently been proposed that the SOC may stabilize a Chern insulator phase 
in the $(001)$ and $(111)$ monolayers of double perovskites~\cite{Cook,AMCook2016} and lead to a topologically nontrivial band structure in $\mathrm{BaTiO_3}$/$\mathrm{Ba_2FeReO_6}$/$\mathrm{BaTiO_3}$ 2D quantum wells~\cite{BFRO}. It should be noted that the tight-binding model for this system with the SOC and tetragonal compression has been investigated~\cite{AMCook2016} in the free fermion level. However, the effect of carrier doping on the magnetism in the presence of the SOC
is still illusive.

In this paper, we study a $(001)$ layer of pure and doped SFMO based on a minimal microscopic model within a large-$S$ expansion. We find that the strong SOC gives rise to a robust nontrivial magnetic state with an electronic structure consisting of four spin-polarized massive or massless Dirac dispersions as well as flat bands.
Based on the spin-wave theory, we demonstrate that such an unusual magnetic state is stable in a large, experimentally relevant range of carrier doping.
In the electron-doped Sr$_{2-x}$La$_x$FeMoO$_6,$ we suggest that the extra electrons would occupy a fully polarized flat band.

\begin{figure}
\centering
\includegraphics[width=0.7\columnwidth]{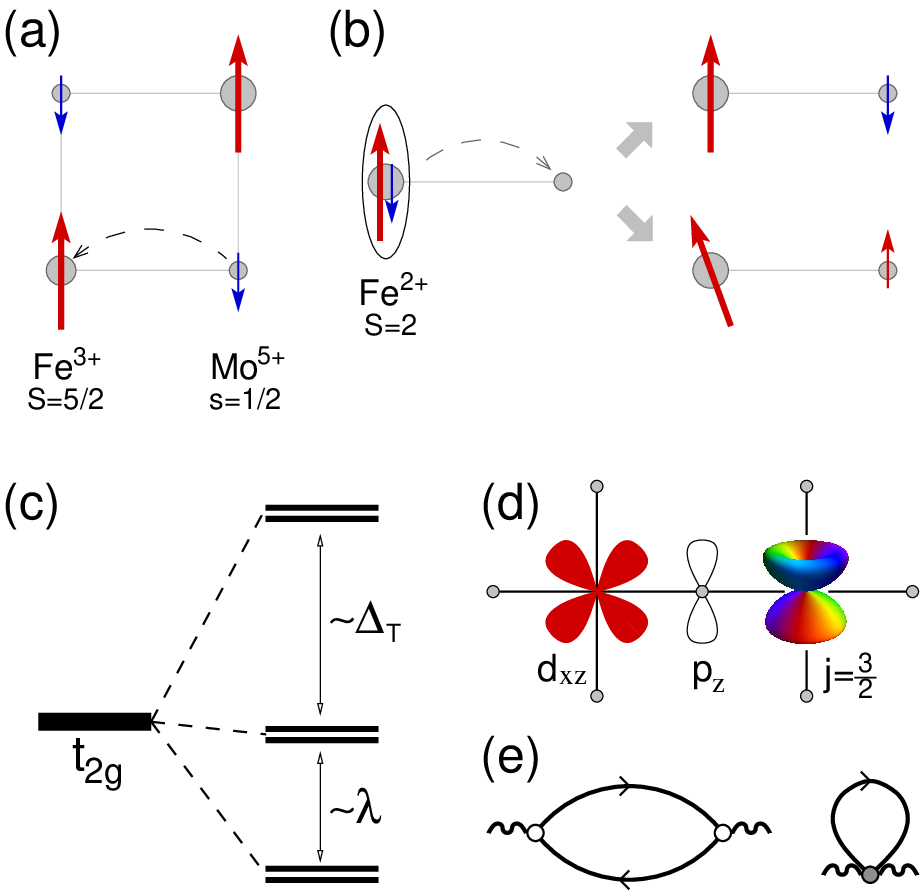}
\caption{(a) Ferrimagnetic order of the Fe$^{3+}~(3d^5)$ $S=5/2$ local moments (long arrows) and the Mo$^{5+}~(4d^1)$ $s=1/2$ states (short arrows) in the ionic picture. Electron transfer from Mo$^{5+}$ to Fe$^{3+}$ is only allowed when its spin is antiparallel to the localized moment, otherwise blocked by the Pauli exclusion principle. (b) When an electron moves from the Fe$^{2+}$ $S=2$ state, an entangled state of the local and carrier spins may leave behind the local moment polarized along the ordered direction (top) or polarized away from it by creating a single magnon state behind (bottom). 
(c) Splitting of the $t_{2g}$ level by a tetragonal crystal field $\Delta_\mathrm{T}$ and the SOC $\lambda$ into three Kramers pairs. (d) Hopping pathway between the Fe $3d_{xz}$ orbital (left) and the Mo $j^z=3/2$ state (right) via the oxygen $2p_z$ orbital. (e) Magnon self-energies in the leading order of the $1/S$ expansion. Solid (wavy) lines represent fermion (magnon) propagators. White (grey) vertices correspond to the coupling between the magnon and the transverse (longitudinal) particle-hole excitations.}
\label{wf}
\end{figure}

\section{Model}
In the (001) layer of double perovskite SFMO, Fe and Mo ions form a checkerboard pattern on a square lattice [see Fig.~\ref{wf}(a)], 
and these metal ions reside inside the oxygen octahedra. In the ionic picture, iron is in the Fe$^{3+}$ valence state with five 3$d$-electrons
coupled by Hund's rule into the high-spin state forming a localized $S=5/2$ moment.
The Mo$^{5+}$ ion has a single 4$d$-electron in the $t_{2g}$ manifold of degenerate $xy$, $xz$, and $yz$ orbitals. The lowest energy coherent charge transfer process takes place when this single electron moves from Mo$^{5+}$ to a neighboring Fe$^{3+}$ through the hybridization between the same $t_{2g}$ states along a given bond. For example, on a bond along the $x$-direction there is a finite overlap either between $xy$ or $xz$ neighboring orbitals with a real amplitude $-t$.

We consider a tetragonal elongation of the oxygen octahedra due to a substrate-induced compressive strain. The corresponding tetragonal crystal field $\Delta_{\rm T}>0$ lifts the threefold $t_{2g}$ orbital degeneracy by stabilizing an orbital doublet of the axial $xz$ and $yz$ 
orbitals and by placing the planar $xy$ orbital at a higher energy.
It should be noted that
for the $xy$ orbital, where the SOC is completely quenched
by the tetragonal distortion, we can repeat the analysis
of the cubic case without the SOC~\cite{Jackeli} to discuss the
stability against doping.
In addition, we include the SOC $\lambda>0$ in Mo$^{5+}$ that
further lifts the degeneracy of
the $xz$ and $yz$ orbitals by stabilizing $j^z=\pm 3/2$ Kramers doublet of the
effective total angular momentum $j=s+l=3/2$ quartet~\cite{SVO}.
Here, $s=1/2$ and $l=1$ are spin and effective angular momentum of a $t_{2g}$
electron, respectively~\cite{Abr70}.
We will not include the SOC for the Fe
3$d$-orbitals because it is much weaker than that for the Mo
4$d$-orbitals, and, moreover, it is fully quenched in the high-spin state of Fe$^{3+}.$
The resulting local energy structure of Mo$^{5+}$ is depicted in Fig.~\ref{wf}(c),
and the explicit forms of $j^z=\pm 3/2$ wave functions are given by
\begin{align}
\ket{j^z\!\!=\!\frac{3}{2}}&=
-\frac{1}{\sqrt{2}}(i\ket{d_{xz\uparrow}}+\ket{d_{yz\uparrow}})\equiv\ket{c_{{\uparrow}}}, \nonumber \\
\ket{j^z\!\!=\!-\frac{3}{2}}&=
-\frac{1}{\sqrt{2}}(i\ket{d_{xz\downarrow}}-\ket{d_{yz\downarrow}})\equiv\ket{c_{{\downarrow}}},
\label{3/2}
\end{align}
and are labeled hereafter by fermionic annihilation operators $c_{\sigma}$ with a pseudospin $\sigma=\,\uparrow,\,\downarrow$.

Here we take the limit of a strong SOC $\lambda$ and a tetragonal field $\Delta_{\rm T}$ compared to the nearest-neighbor (NN) hopping amplitude $|t|$. Projecting the $t_{2g}$ orbitals onto the lowest energy states \eqref{3/2}, we have obtained a low-energy Hamiltonian for a charge transfer between NN Fe and Mo ions [see Fig.~\ref{wf}(d)], as follows.
\begin{align}
\mathcal{H}_t= &\frac{t}{\sqrt{2}}{\Biggl[}\sum_{\langle ij\rangle \in x,\,\sigma} id_{i,xz\sigma}^\dagger c_{j,\sigma}
+\sum_{\langle ij\rangle \in y,\,\sigma}\sigma d_{i,yz\sigma}^\dagger c_{j,\sigma}
+\textrm{h.c.}{\Biggr]} \nonumber \\
&+\Delta{\Bigl[} \sum_i n^{(d)}_{i }-\sum_j n^{(c)}_{j}{\Bigr]}, \label{eqht}
\end{align}
where $i(j)$ labels Fe(Mo) ions, $\langle ij\rangle \in x(y)$ refers to each NN bond along the $x(y)$-direction, $\sigma=\,\uparrow,\,\downarrow\,=\pm 1$ stands for a spin index, $2\Delta$ is a charge transfer gap between Mo$^{5+}$ and Fe$^{3+}$, the number operators $n^{(d)}_{i}$ and $n^{(c)}_{j}$ measure carrier density $d_i^\dagger d_i$ and $c_j^\dagger c_j$,
respectively. In the undoped SFMO, there is one carrier $n=n^{(d)}+n^{(c)}=1$ per formula unit, ignoring the localized half-filled Fe $d$-shell.
The SOC manifests itself in the spin-dependent hopping in~Eq.~\eqref{eqht} that explicitly breaks the original $SU(2)$ symmetry.
Hereafter, we set $t=1$ as the energy scale for simplicity.

When an itinerant electron visits the Fe$^{3+}$ ion with a core spin $S=5/2$, the resulting total spin $\mathcal{S}$ of Fe$^{2+}$ could, in principle, take one of the two possible values $\mathcal{S}=2$ and $\mathcal{S}=3.$ However, the maximum allowed spin quantum number for six electrons in a $d$-shell is $\mathcal{S}=2$. The unphysical $\mathcal{S}=3$ states appear because the local and itinerant spin operators on the Fe site are treated as independent variables. In order to project the enlarged Hilbert space onto the physical one, we supplement the hopping Hamiltonian~\eqref{eqht} by a local antiferromagnetic (AF) coupling $J\to \infty$ between the local and itinerant spins~\cite{Jackeli}. The total Hamiltonian then becomes $\mathcal{H}=\mathcal{H}_t+\mathcal{H}_{\mathcal{S}}$, where
\begin{equation}
		\mathcal{H}_{\mathcal{S}} = J\sum_i \left[\vec{S}_i\cdot \vec{s}_i+\frac{S+1}{2} n^{(d)}_{i} \right].
\label{eqfe}
\end{equation}
The sum is taken over every Fe site $i$, and $\vec{S}_i$ and $\vec{s}_i$ are operators for the local and itinerant spins, respectively.
\begin{figure}
\centering
\includegraphics[width=0.8\columnwidth]{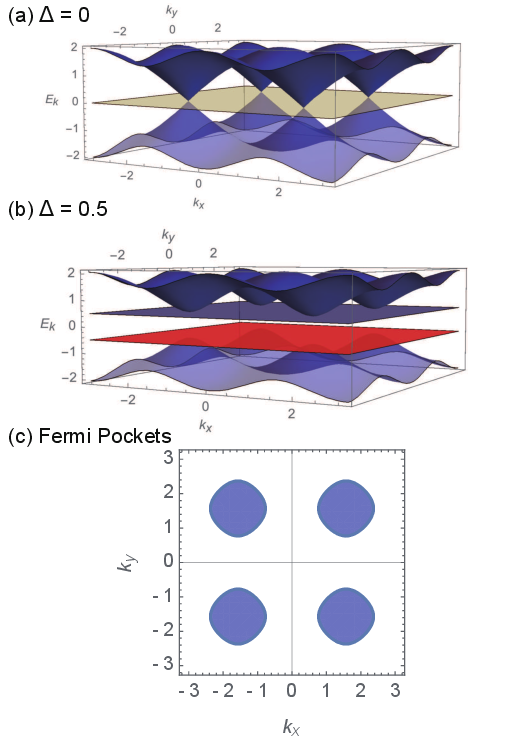} 
\caption{Band structure of the single-electron Hamiltonian~\eqref{bands} for (a) gapless $\Delta=0$ and (b) gaped $\Delta\neq0$ cases 
(see the text for the nature of the bands). (c) Fermi pockets of the lowest band for a carrier density $n=0.75$ per formula unit with $\Delta=0$.}
\label{band}
\end{figure}

\section{Ferrimagnetic ground state}
The model defined by~Eqs.~\eqref{eqht} and \eqref{eqfe} is one version of canonical double exchange (DE) problems with an infinite exchange coupling between the local and itinerant spins. Similarly to the DE, a maximum kinetic energy gain is achieved when the local moments align ferromagnetically (FM) and, in the present case, antiparallel to the itinerant spins, giving rise to an FiM state.

We consider the FiM order along the tetragonal symmetry $z$-axis and discuss later its stability within the large-$S$ spin-wave theory. We introduce fermionic operators $D_{\downarrow(\uparrow)}$ for the carriers on the Fe sites with their spins quantized along the local moments. This representation diagonalizes the spin part of the Hamiltonian~Eq.~\eqref{eqfe} and projects out the fermionic states $D_{\uparrow}$ corresponding to the unphysical states with $\mathcal{S}=3$ (see Ref.~\cite{Jackeli} for details). The $d$-operators in Eq.~\eqref{eqht} in terms of the new ones are expressed as $d_{xz(yz)\downarrow}=D_{xz(yz)\downarrow}[1-b^\dagger b/(4S)]$ and $d_{xz(yz)\uparrow}=-D_{xz(yz)\downarrow}b^\dagger/\sqrt{2S}$, where $b$ is
a bosonic annihilation operator for a single magnon state. This is created when a spin-down electron moves away from Fe$^{2+}$, which is in
the entangled $\mathcal{S}=2$ state of the local and carrier spins, leaving an Fe$^{3+}$ local moment in the $S^z=S-1=3/2$ single magnon state which is tilted away from the fully polarized $S=S^z=5/2$ state [see Fig.~\ref{wf}(b)]. This representation provides the correct matrix elements of fermionic operators within the eigenstates of the allowed total spin, $S=5/2$ and $\mathcal{S}=2$, in the perturbative level and retains a quantum nature of the local moments.
The large-$S$ expansion is justified by the smallness of $1/S=2/5.$
Indeed, the next-order contribution of $1/S^2$ is irrelevant for a small doping level
in the previous calculation~\cite{Jackeli}, and thus we expect that the expansion until $1/S$ is enough
in our case.  The approximation to include a strong SOC in Mo and no SOC in Fe can be
justified easily by the fact that SOC is completely quenched in spherical half-filled
$d$-orbitals of Fe$^{3+},$ while SOC is active in the $t_{2g}$-orbitals of Mo~\cite{Yamada2018}.

Inserting the above representation into~Eq.~\eqref{eqht}, we get
$\mathcal{H}=\mathcal{H}_0+\mathcal{H}_1+\mathcal{H}_2$, where $\mathcal{H}_0$ is a single-particle part and $\mathcal{H}_{1(2)}\sim1/\sqrt{S}~(1/S)$ refers
to the coupling of the magnons with the transverse (longitudinal) particle-hole pairs.
 
Diagonalizing the noninteracting $\mathcal{H}_0$ part, we get the following expression in the momentum space.
\begin{equation}
\mathcal{H}_0= \sum_k {\Bigr\{}\Delta[\alpha_{k\downarrow}^\dagger \alpha_{k\downarrow}-c_{k\uparrow}^\dagger c_{k\uparrow}]+E_k[\beta_{k\downarrow}^\dagger \beta_{k\downarrow}-\gamma_{k\downarrow}^\dagger \gamma_{k\downarrow}]{\Bigl\}},
\label{bands}
\end{equation}
where $E_k=\sqrt{\Delta^2+2(\cos^2 k_x+\cos^2 k_y)}$ and the eigenstates $\alpha_{k\downarrow},\,\beta_{k\downarrow},\,\gamma_{k\downarrow},$ and $c_{k\uparrow}$ have been obtained by a unitary transformation.
The band structure~\eqref{bands} is composed of four bands and is shown in Fig.~\ref{band}(a)-(b). The two flat $\alpha_\downarrow$ and $c_\uparrow$ bands correspond to a nonbonding state composed of the $d_{xz\downarrow}$ and $d_{yz\downarrow}$ orbitals of Fe and a localized $j^z=3/2$ state of Mo, respectively. The dispersive antibonding $\beta_\downarrow$ and bonding $\gamma_\downarrow$ bands are made of spin-down states of Mo and Fe. The next-nearest-neighbor (NNN) hopping between the same Fe (or Mo) ions, not considered here, might in principle give a finite dispersion to the flat bands. However, the corresponding hopping is between the $d_{xz}$ and $d_{yz}$ orbitals, and is extremely small ($\sim$ few meV)~\cite{Cook}. Moreover, it exactly vanishes when projected onto the complex wave functions of the Mo $j^z=\pm3/2$ states due to a destructive quantum interference.
Notice that the robust half metallicity and the flat band owe the half-filled $d$-orbitals of
Fe$^{3+},$ which is different from Cr$^{3+}$ in Ref.~\cite{AMCook2016}.

\begin{figure}
\centering
\includegraphics[width=0.7\columnwidth]{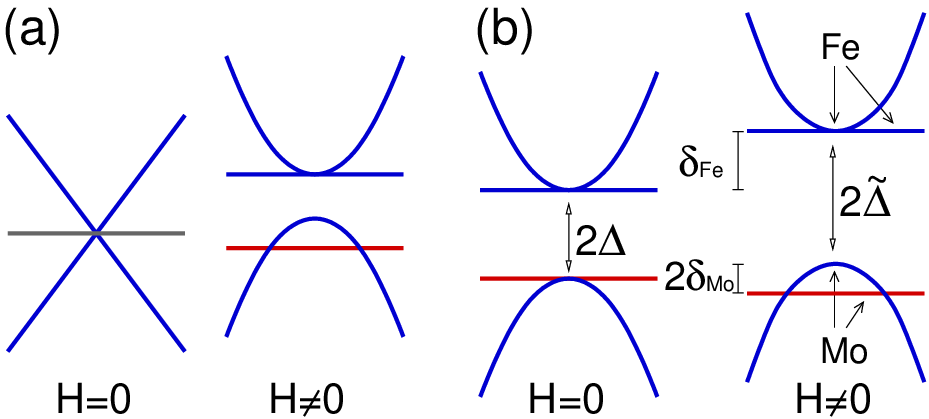} 
\caption{Schematic view of the band structure renormalisation induced by an applied magnetic field ($H$) for (a) gapless $\Delta=0$ and (b) gapped $\Delta\neq0$ cases. Spin-up, spin-down, and spin-unpolarized states are marked by red, blue, and grey colors, respectively. The spin-down (spin-up) flat band has a purely Fe (Mo) character.
The higher-(lower-) energy dispersive band at its minima (maxima) is solely made of the Fe (Mo) states. In both cases, (a) and
(b), an applied field shifts up the higher-energy Fe bands by $\delta_{\rm Fe}=g_{\rm Fe}\mu_BH/2$ and splits
the lower-energy Mo flat and dispersive bands by $2\delta_{\rm Mo}=g_{\rm Mo}\mu_BH$.
Here, $g_{\rm Fe}\simeq 2$ and $g_{\rm Mo}\ll 1$ are the corresponding
$g$-factors~\cite{g}, and $2\tilde{\Delta}=2\Delta+(g_{\rm Fe}-g_{\rm Mo})\mu_BH/2$ is a renormalized mass gap.}
\label{field}
\end{figure}

We now discuss the effects of a uniform external magnetic field $H$ on $\mathcal{H}_0$. This just splits the four bands without hybridization because $\mathcal{H}_0$ conserves the $z$-component of the real spins. While the external field is uniform, the Zeeman splitting $g\mu_B H$ of the itinerant electrons on the Fe and Mo ions become staggered due to the difference in the $g$-factors.
As shown in Fig.~\ref{field}, this would allow us to control the mass of Dirac dispersions
and also to dope the flat band just above the Fermi energy.

The flat bands, which are already fully spin-polarized, are different from the unpolarized ones, such as the ones in the $(110)$ thin films of STO~\cite{Wang} or in (metal)-organic systems~\cite{Garnica,Yamada}, supporting the flat-band ferromagnetism~\cite{Lieb,Mielke,Tasaki}.
Therefore, in electron-doped Sr$_{2-x}$La$_x$FeMoO$_6$ or SFMO under a strong magnetic field, where extra electrons occupy the nondispersive Mo band, we anticipate other types of instabilities, such as Wigner crystallization~\cite{Wigner} or various types of complex charge ordered patterns, as well as the formation of self-trapped polaronic states of minority spins at the Mo sites. As confirmed in the following part, the FiM state is stable in a wide carrier doping range of the electron doping, and, consequently, the minority-spin flat Mo band can indeed be electron-doped.

\section{Spin-wave spectrum}
We now analyze the stability of the FiM order state postulated above. To this end,
we derive a spin-wave excitation spectrum from the magnon Green function $G_{q,\omega}=1/[\omega -\Sigma_{q,\omega}]$ evaluated within the leading order of the large-$S$ expansion.
First, we note that in the classical $S\to\infty$ limit, the magnons are localized, and they become dispersive only due to quantum corrections.
The corresponding magnon self-energy corrections ($\Sigma_{q,\omega}\sim 1/S$), shown in Fig.~\ref{wf}(c), stem from magnon interactions with propagating transverse and longitudinal particle-hole excitations. Their expressions are quite lengthy and given in Appendix. We find that a coherent spin-wave mode emerges below the Stoner continuum with the following dispersion relation in the low-energy limit.
\begin{align}
\omega_q &=4J_1 S[1+\Gamma_{1q}]+8J_2 S[1-\Gamma_{2q}],
\label{SW}
\end{align}
where $\Gamma_{1q} =\cos q_x \cos q_y$, $\Gamma_{2q}=(\cos^2 q_x +\cos^2 q_y)/2$, $J_1$ and $J_2$ are the
carrier induced exchange couplings between the NN and NNN local moments, respectively. They depend only on the carrier density and the charge transfer gap $\Delta$.
We note that the ferromagnetic state of the localized moments 
is the exact ground state of the effective anisotropic spin-only Hamiltonian 
[see Appendix~\ref{hidden}].

The spectrum~\eqref{SW} is gapless at $q=(\pi,0)$ and at the symmetry-related points, which is very surprising because (i) the model defined by
Eqs.~\eqref{eqht}~and~\eqref{eqfe} does not host any apparent continuous spin-rotation symmetry, and (ii) the gapless points are away from the FM Bragg point $q=(0,0)$. Actually, the model has a hidden $SU(2)$ symmetry that can be uncovered by a gauge transformation.

\begin{figure}
\centering
\includegraphics[width=\columnwidth]{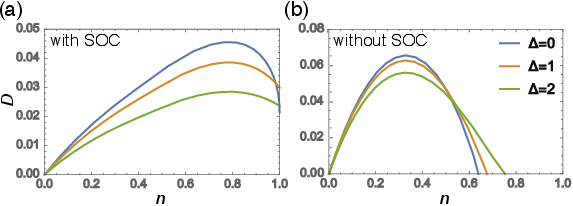} 
\caption{(a) Spin stiffness $\mathcal{D}$ of the FiM ordered state
versus carrier density $n$ for $0<n<1$ at various values of the gap $\Delta$. For $1<n<2$ (not shown), the spin stiffness remains nearly constant. (b) For comparison, same as (a) but without SOC~\cite{Jackeli}.}
\label{spin}
\end{figure}

For the spectrum~\eqref{SW}, the spin stiffness of the FiM ordered state is given by $\mathcal{D}=2J_1 S+4J_2 S$. Shown in Fig.~\ref{spin}(a) is the dependence of $\mathcal{D}$ on the carrier density $n$ for $0<n<1$ at various values of the band gap $\Delta$.
In the range $1<n<2$, $\mathcal{D}$ remains constant for $\Delta=0$. This is because the added electron carriers occupy the unpolarized flat band, and no additional potential or kinetic energy is gained.
For $\Delta\neq0$, $\mathcal{D}$ becomes very weakly renormalized ($\sim$ few percent)
as carriers occupy the minority spin flat band.
For comparison, in Fig.~\ref{spin}(b) we plot the spin stiffness obtained at a zero SOC~\cite{Jackeli}. $\mathcal{D}$ vanishes at some critical doping, signaling the instability of the
FiM order. Without SOC the FiM ground state cannot be stabilized at $n=1$ or $n$ slightly larger than 1.

Thus, a strong SOC extends the stability window of the FiM order to the experimentally accessible electron doping range.
The reason behind the extended stability is the SOC-induced electronic band structure reconstruction which transforms a large Fermi surface centered around the $q=(0,0)$ to four small Fermi pockets around $(\pm\frac{\pi}{2},\pm\frac{\pi}{2})$, as shown in Fig.~\ref{band}(c), allowing more kinetic energy gain with an increasing carrier density. Indeed, as seen experimentally~\cite{Sanchez}, electron-doped Sr$_{2-x}$La$_x$FeMoO$_6$ thin films do exhibit the stable FiM order in a wide range of doping as correctly shown here by the model with the SOC.
We note that the disorder effect in Sr$_{2-x}$La$_x$FeMoO$_6$ does not
affect the flat band as electrons in the flat band are already localized.

In conclusion, we have proven the stability of the FiM ground state of
SFMO thin films against doping within a perturbative analysis. We discovered that
the SOC plays a critical role in the enhancement of the stability and results in a phase
with an unusual band structure including Dirac dispersions and flat bands.
We anticipate that this gives rise to
interesting collective behaviors, such as Wigner crystallization~\cite{Wigner}.
The disorder effect in a Wigner crystal has also
been discussed in Ref.~\cite{Chitra2001} as a pinned Wigner crystal.

The above proposed non-trivial band structure in this quasi-2D
system can be directly verified through angle resolved photoemission spectroscopy (ARPES),
and we believe that our theoretical findings will motivate future ARPES
experiments.

\begin{acknowledgments}
		We wish to acknowledge A.F.~Bangura, J.A.N.~Bruin, A.M.~Cook, D.~Hirai, I.H.~Inoue, J.~Lee, J.~Romh\'anyi, A.~Paramekanti, D.D.~Sarma, S.~Tsuneyuki, T.~Vekua, M.~Oshikawa, and, especially, H.~Takagi for illuminating discussions. We thank A.~Subedi for the critical reading of the manuscript and comments.
M.G.Y. acknowledges the opportunity of his internship and a kind hospitality at the Quantum Materials Department, Max Planck Institute for Solid State Research, Stuttgart, Germany, where the major part of this work was done. A part of this work was completed at MPI-PKS international workshop ``Topological Phenomena in Novel Quantum Matter'', and at KITP, UCSB, supported by the US National Science Foundation under Grant Nos. NSF PHY11-25915, PHY-1748958.
This work was also supported by JST CREST Grant Number JPMJCR19T5, Japan,
by JSPS KAKENHI Grant Numbers JP15H02113, JP17J05736,
and by JSPS Strategic International Networks Program No. R2604 ``TopoNet''.
M.G.Y. is supported by the Materials Education program for the future leaders in Research, Industry, and Technology (MERIT) and by JSPS. We also acknowledge the support of the Max-Planck-UBC-UTokyo Centre for Quantum Materials.
\end{acknowledgments}

\appendix

\section{Diagonalization of the free-fermion part}

The eigenstates of the tight-binding model are obtained by a unitary transformation as
\begin{align}
\alpha_{k\downarrow}&=-iv_k D_{xz,k\downarrow}+u_k D_{yz,k\downarrow}, \nonumber \\
\beta_{k\downarrow}&=\tilde{u}_k \tilde{D}_{k\downarrow}-\tilde{v}_k e^{ik_x}c_{k\downarrow}, \nonumber \\
\gamma_{k\downarrow}&=\tilde{v}_k \tilde{D}_{k\downarrow}+\tilde{u}_k e^{ik_x}c_{k\downarrow}, \nonumber \\
\tilde{D}_{k\downarrow}&=iu_k D_{xz,k\downarrow}+v_k D_{yz,k\downarrow},
\end{align}
with the following factors:
$u_k=T_{x,k}/\varepsilon_k$, $v_k=T_{y,k}/\varepsilon_k$,
$\tilde{u}_k=\sqrt{(1+\Delta/E_k)/2}$, $\tilde{v}_k=\sqrt{(1-\Delta/E_k)/2}$, and
$T_{x(y),k}=\sqrt{2}\cos k_{x(y)}$,
where $E_k=\sqrt{\varepsilon_k^2+\Delta^2}$ and $\varepsilon_k=\sqrt{T_{x,k}^2+T_{y,k}^2}$.

\section{Magnon-fermion interaction}

The interaction between the magnons and the transverse and longitudinal components of particle-hole excitations is
described by the following $\mathcal{H}_1\sim 1/\sqrt{S}$ and $\mathcal{H}_2\sim 1/S$ terms, respectively:

\begin{align}
\mathcal{H}_1 &= \frac{1}{\sqrt{2SN}} \sum_{k,q} \{ c_{k-q\uparrow}^\dagger [L_{k,q}\alpha_{k\downarrow}+M_{k,q}\beta_{k\downarrow}+N_{k,q}\gamma_{k\downarrow}] b_q^\dagger \nonumber \\
&\times e^{i(q_x-k_x)}+\textrm{h.c.}\}, \\
\mathcal{H}_2 &= \frac{1}{4SN}\sum_{k,p,q}b_q^\dagger b_{q+k-p}\{P_{k,p}^{\alpha\beta}\alpha_{k\downarrow}^\dagger\beta_{p\downarrow}+P_{p,k}^{\alpha\beta}\beta_{k\downarrow}^\dagger\alpha_{p\downarrow} \nonumber \\
&+P_{k,p}^{\alpha\gamma}\alpha_{k\downarrow}^\dagger\gamma_{p\downarrow}+P_{p,k}^{\alpha\gamma}\gamma_{k\downarrow}^\dagger\alpha_{p\downarrow}
+P_{k,p}^{\beta\beta}\beta_{k\downarrow}^\dagger\beta_{p\downarrow} \nonumber \\
&+P_{k,p}^{\beta\gamma}\beta_{k\downarrow}^\dagger\gamma_{p\downarrow}
+P_{p,k}^{\beta\gamma}\gamma_{k\downarrow}^\dagger\beta_{p\downarrow}+P_{k,p}^{\gamma\gamma}\gamma_{k\downarrow}^\dagger\gamma_{p\downarrow}\},
\end{align}
with interaction vertices
\begin{align}
L_{k,q}&=-T_{x,k-q}v_k-T_{y,k-q}u_k, \nonumber \\
M_{k,q}&=T_{x,k-q}u_k \tilde{u}_k-T_{y,k-q}v_k \tilde{u}_k, \nonumber \\
N_{k,q}&=T_{x,k-q}u_k \tilde{v}_k-T_{y,k-q}v_k \tilde{v}_k, \nonumber \\
P_{k,p}^{\alpha\beta}&=T_{x,p}v_k\tilde{v}_p-T_{y,p}u_k\tilde{v}_p, \nonumber \\
P_{k,p}^{\alpha\gamma}&=-T_{x,p}v_k\tilde{u}_p+T_{y,p}u_k\tilde{u}_p, \nonumber \\
P_{k,p}^{\beta\beta}&=-[(T_{x,p}u_k+T_{y,p}v_k)\tilde{u}_k\tilde{v}_p+(k\leftrightarrow p)], \nonumber \\
P_{k,p}^{\beta\gamma}&=(T_{x,p}u_k+T_{y,p}v_k)\tilde{u}_k\tilde{u}_p-(k\leftrightarrow p,\,\tilde{u}\leftrightarrow \tilde{v}), \nonumber \\
P_{k,p}^{\gamma\gamma}&=(T_{x,p}u_k+T_{y,p}v_k)\tilde{v}_k\tilde{u}_p+(k\leftrightarrow p),
\end{align}
where $(k\leftrightarrow p)$ means the corresponding term with $k$ exchanged for $p$.

\section{Details of the spin-wave spectrum}

The spin-wave spectrum is derived from the poles of the magnon Green function $G_{q,\omega}=1/[\omega -\Sigma_{q,\omega}]$,
where the magnon self energy $\Sigma_{q,\omega}$ is composed of two parts within the leading order of $1/S$ expansion.
They are diagrammatically depicted in Fig.~1(c) in the main text and can be grouped into the following algebraic form when $0<n<2$,
\begin{align}
 \Sigma_{q,\omega}&=\frac{1}{2SN}\sum_k{\Bigr\{}|L_{k,q}|^2\frac{n_{c,k-q\uparrow}-n_{\alpha,k\downarrow}}{\omega-2\Delta} \nonumber \\
 &+|M_{k,q}|^2\frac{n_{c,k-q\uparrow}-n_{\beta,k\downarrow}}{\omega-E_k-\Delta}
 +|N_{k,q}|^2\frac{n_{c,k-q\uparrow}-n_{\gamma,k\downarrow}}{\omega+E_k-\Delta}{\Bigr\}} \nonumber \\
 &+\frac{1}{4SN}\sum_k P_{k,k}^{\gamma\gamma} n_{\gamma,k\downarrow}.
 \label{green}
\end{align}
Here, $n_{m,k\sigma}$ is the occupation number of the $m_{k\sigma}$ state ($m=\alpha,\beta,\gamma,c$) depending on the total carrier density and $N$ is the number of unit cells.

From the poles of $G$, we got a Stoner continuum and a gapless mode of the spin wave in the order of $1/S$.
Assuming $|\omega|$ to be small, the obtained low-energy spin wave has a form of a localized spin model with
anisotropic nearest-neighbor (NN) and isotropic next-nearest-neighbor (NNN) exchange couplings, $J_1$ and $J_2$, respectively [see the next section].
For the carrier density $n\leq1$ only the lowest dispersive mode is occupied and exchange couplings, $J_1$ and $J_2$, are given by
\begin{align}
J_1&=\frac{1}{8S^2N}\sum_k T_{x,k}^2 T_{y,k}^2\frac{n_{\gamma,k\downarrow}}{E_k \varepsilon_k^2}, \nonumber \\
J_2&=\frac{1}{8S^2N} \sum_k \cos(2k_x) T_{x,k}^2  \frac{n_{\gamma,k\downarrow}}{E_k \varepsilon_k^2}.
\end{align}

For the filling $n>1$, we need to treat the case $\Delta=0$ and the case $\Delta>0$ separately. When $\Delta=0,$ the band structure consists of Dirac-like dispersions and two degenerate flat bands $\alpha$ and $c$, with opposite spin polarization [see Fig.~2(a) in the main text]. Due to the degeneracy $n_{c,k-q\uparrow}=n_{\alpha,k\downarrow}$, the first Lindhard function in Eq.~\eqref{green} vanishes, and we can then take the approximation $\omega/E_k \to 0$. We can easily find that the sum of the second and third Lindhard functions also vanish other than the term including $n_{\gamma,k\downarrow}$ in this low-energy limit, so the spectrum of the low-energy mode (the pole of the Green function) does not change for $1<n<2$. This is because the flat bands cannot earn any kinetic or potential energy when $\Delta=0$ while ferrimagnetism is stabilized here in order to earn the total energy of electrons.

Next, we consider the case $\Delta>0$ with massive Dirac dispersions. In the limit $\omega/\Delta \to 0,$ the sum of the second and third Lindhard functions in Eq.~\eqref{green} vanishes other than the term including $n_{\gamma,k\downarrow}$ again. We find the overall 
renormalisation of the low-energy mode for $n=1+x$ as $\omega_q(x)=\omega_q+\delta\omega_q(x)$, with
\begin{align}
		\delta\omega_q(x) &= -\frac{1}{2SN}\sum_k|L_{k.q}|^2\frac{n_{c,k-q\uparrow}}{2\Delta} \nonumber \\
        &=x\{4j_1 S[1+\Gamma_{1q}]+8j_2S[1-\Gamma_{2q}]\}.
\end{align}
\begin{align}
j_1&=-\frac{1}{2S^2N}\sum_k \frac{\cos^2k_x\cos^2k_y}{\Delta\varepsilon_k^2}\simeq -\frac{0.045}{S^2\Delta}, \nonumber \\
j_2&=\frac{1}{4S^2N}\sum_k \frac{\cos2k_x\cos^2k_y}{\Delta\varepsilon_k^2}\simeq -\frac{0.017}{S^2\Delta}.
\end{align}
The corrections appear to be negligibly small, amounting to only a few percent change of exchange couplings for electron doping in the physically accessible range $1<n<2$.

\section{Hidden $SU(2)$ symmetry}\label{hidden}

In order to uncover the origin of the gapless mode discussed above, we note that the obtained spin-wave spectrum,~Eq.~(5) in the main text, is exactly the same as that of the following spin-only model on the square lattice of iron sites for local spin moments:
 \begin{equation}
        \mathcal{H}_\textrm{spin}=J_1 \sum_{\langle ij \rangle}[S_i^x S_j^x+S_i^y S_j^y-S_i^z S_j^z]-J_2 \sum_{\langle\!\langle ij \rangle\!\rangle} \vec{S}_i \cdot \vec{S}_j, \label{spinmodel}
\end{equation}
with anisotropic NN ($J_1$) and isotropic NNN ($J_2$) couplings. It is straightforward to verify that the model~\eqref{spinmodel}
hosts a hidden spin-rotation symmetry and can be mapped to the isotropic ferromagnetic Heisenberg model by rotating spins sitting on the odd rows along the $x$-direction (corresponding to one set of staggered sublattice of the square lattice of iron sites) around the $z$-axis by an angle $\pi$ $(S_i^x\to-S_i^x,~S_i^y\to-S_i^y,~S_i^z=S_i^z)$. This site-dependent spin rotation also shifts $q=(\pi,0)\to (0,0)$. This explains the origin and the location of the gapless Nambu-Goldstone mode accompanied by the spontaneous symmetry breaking of this hidden $SU(2)$ symmetry. Indeed, the same spin rotation for fermionic operators ($d_{ixz(yz)\sigma}\to\sigma d_{ixz(yz)\sigma},~c_{i\sigma}\to\sigma c_{i\sigma}$) at every second row together with local moments, maps the hopping Hamiltonian [see Eq.~(2) in the main text] to the $SU(2)$-invariant form and leaves the local one [Eq.~(3)] unchanged. However, we point out that this emergent spin-rotation symmetry is only approximate in reality and exists only in the extreme $\lambda\to\infty$ limit considered here, and thus the ferrimagnetic order should be pinned along the $z$-direction.

\bibliography{paper}

\begin{thebibliography}{44}%
\makeatletter
\providecommand \@ifxundefined [1]{%
 \@ifx{#1\undefined}
}%
\providecommand \@ifnum [1]{%
 \ifnum #1\expandafter \@firstoftwo
 \else \expandafter \@secondoftwo
 \fi
}%
\providecommand \@ifx [1]{%
 \ifx #1\expandafter \@firstoftwo
 \else \expandafter \@secondoftwo
 \fi
}%
\providecommand \natexlab [1]{#1}%
\providecommand \enquote  [1]{``#1''}%
\providecommand \bibnamefont  [1]{#1}%
\providecommand \bibfnamefont [1]{#1}%
\providecommand \citenamefont [1]{#1}%
\providecommand \href@noop [0]{\@secondoftwo}%
\providecommand \href [0]{\begingroup \@sanitize@url \@href}%
\providecommand \@href[1]{\@@startlink{#1}\@@href}%
\providecommand \@@href[1]{\endgroup#1\@@endlink}%
\providecommand \@sanitize@url [0]{\catcode `\\12\catcode `\$12\catcode
  `\&12\catcode `\#12\catcode `\^12\catcode `\_12\catcode `\%12\relax}%
\providecommand \@@startlink[1]{}%
\providecommand \@@endlink[0]{}%
\providecommand \url  [0]{\begingroup\@sanitize@url \@url }%
\providecommand \@url [1]{\endgroup\@href {#1}{\urlprefix }}%
\providecommand \urlprefix  [0]{URL }%
\providecommand \Eprint [0]{\href }%
\providecommand \doibase [0]{http://dx.doi.org/}%
\providecommand \selectlanguage [0]{\@gobble}%
\providecommand \bibinfo  [0]{\@secondoftwo}%
\providecommand \bibfield  [0]{\@secondoftwo}%
\providecommand \translation [1]{[#1]}%
\providecommand \BibitemOpen [0]{}%
\providecommand \bibitemStop [0]{}%
\providecommand \bibitemNoStop [0]{.\EOS\space}%
\providecommand \EOS [0]{\spacefactor3000\relax}%
\providecommand \BibitemShut  [1]{\csname bibitem#1\endcsname}%
\let\auto@bib@innerbib\@empty
\bibitem [{\citenamefont {Boschker}\ and\ \citenamefont
  {Mannhart}(2017)}]{Bosh2017}%
  \BibitemOpen
  \bibfield  {author} {\bibinfo {author} {\bibfnamefont {H.}~\bibnamefont
  {Boschker}}\ and\ \bibinfo {author} {\bibfnamefont {J.}~\bibnamefont
  {Mannhart}},\ }\href {\doibase 10.1146/annurev-conmatphys-031016-025404}
  {\bibfield  {journal} {\bibinfo  {journal} {Annu. Rev. Condens. Matter
  Phys.}\ }\textbf {\bibinfo {volume} {8}},\ \bibinfo {pages} {145} (\bibinfo
  {year} {2017})}\BibitemShut {NoStop}%
\bibitem [{\citenamefont {Hwang}\ \emph {et~al.}(2012)\citenamefont {Hwang},
  \citenamefont {Iwasa}, \citenamefont {Kawasaki}, \citenamefont {Keimer},
  \citenamefont {Nagaosa},\ and\ \citenamefont {Tokura}}]{Hwan2012}%
  \BibitemOpen
  \bibfield  {author} {\bibinfo {author} {\bibfnamefont {H.~Y.}\ \bibnamefont
  {Hwang}}, \bibinfo {author} {\bibfnamefont {Y.}~\bibnamefont {Iwasa}},
  \bibinfo {author} {\bibfnamefont {M.}~\bibnamefont {Kawasaki}}, \bibinfo
  {author} {\bibfnamefont {B.}~\bibnamefont {Keimer}}, \bibinfo {author}
  {\bibfnamefont {N.}~\bibnamefont {Nagaosa}}, \ and\ \bibinfo {author}
  {\bibfnamefont {Y.}~\bibnamefont {Tokura}},\ }\href@noop {} {\bibfield
  {journal} {\bibinfo  {journal} {Nat. Mater.}\ }\textbf {\bibinfo {volume}
  {11}},\ \bibinfo {pages} {103} (\bibinfo {year} {2012})}\BibitemShut
  {NoStop}%
\bibitem [{\citenamefont {Takagi}\ and\ \citenamefont
  {Hwang}(2010)}]{Taka2010}%
  \BibitemOpen
  \bibfield  {author} {\bibinfo {author} {\bibfnamefont {H.}~\bibnamefont
  {Takagi}}\ and\ \bibinfo {author} {\bibfnamefont {H.~Y.}\ \bibnamefont
  {Hwang}},\ }\href {\doibase 10.1126/science.1182541} {\bibfield  {journal}
  {\bibinfo  {journal} {Science}\ }\textbf {\bibinfo {volume} {327}},\ \bibinfo
  {pages} {1601} (\bibinfo {year} {2010})}\BibitemShut {NoStop}%
\bibitem [{\citenamefont {Schlom}\ \emph {et~al.}(2008)\citenamefont {Schlom},
  \citenamefont {Chen}, \citenamefont {Pan}, \citenamefont {Schmehl},\ and\
  \citenamefont {Zurbuchen}}]{JACE:JACE02556}%
  \BibitemOpen
  \bibfield  {author} {\bibinfo {author} {\bibfnamefont {D.~G.}\ \bibnamefont
  {Schlom}}, \bibinfo {author} {\bibfnamefont {L.-Q.}\ \bibnamefont {Chen}},
  \bibinfo {author} {\bibfnamefont {X.}~\bibnamefont {Pan}}, \bibinfo {author}
  {\bibfnamefont {A.}~\bibnamefont {Schmehl}}, \ and\ \bibinfo {author}
  {\bibfnamefont {M.~A.}\ \bibnamefont {Zurbuchen}},\ }\href@noop {} {\bibfield
   {journal} {\bibinfo  {journal} {J. Am. Ceram. Soc.}\ }\textbf {\bibinfo
  {volume} {91}},\ \bibinfo {pages} {2429} (\bibinfo {year}
  {2008})}\BibitemShut {NoStop}%
\bibitem [{\citenamefont {Chen}\ \emph {et~al.}(2016)\citenamefont {Chen},
  \citenamefont {Hu}, \citenamefont {Lu}, \citenamefont {Yang}, \citenamefont
  {Zhang}, \citenamefont {Li}, \citenamefont {Ahmed}, \citenamefont {Enriquez},
  \citenamefont {Weigand}, \citenamefont {Su}, \citenamefont {Wang},
  \citenamefont {Zhu}, \citenamefont {MacManus-Driscoll}, \citenamefont {Chen},
  \citenamefont {Yarotski},\ and\ \citenamefont {Jia}}]{Chene1600245}%
  \BibitemOpen
  \bibfield  {author} {\bibinfo {author} {\bibfnamefont {A.}~\bibnamefont
  {Chen}}, \bibinfo {author} {\bibfnamefont {J.-M.}\ \bibnamefont {Hu}},
  \bibinfo {author} {\bibfnamefont {P.}~\bibnamefont {Lu}}, \bibinfo {author}
  {\bibfnamefont {T.}~\bibnamefont {Yang}}, \bibinfo {author} {\bibfnamefont
  {W.}~\bibnamefont {Zhang}}, \bibinfo {author} {\bibfnamefont
  {L.}~\bibnamefont {Li}}, \bibinfo {author} {\bibfnamefont {T.}~\bibnamefont
  {Ahmed}}, \bibinfo {author} {\bibfnamefont {E.}~\bibnamefont {Enriquez}},
  \bibinfo {author} {\bibfnamefont {M.}~\bibnamefont {Weigand}}, \bibinfo
  {author} {\bibfnamefont {Q.}~\bibnamefont {Su}}, \bibinfo {author}
  {\bibfnamefont {H.}~\bibnamefont {Wang}}, \bibinfo {author} {\bibfnamefont
  {J.-X.}\ \bibnamefont {Zhu}}, \bibinfo {author} {\bibfnamefont {J.~L.}\
  \bibnamefont {MacManus-Driscoll}}, \bibinfo {author} {\bibfnamefont {L.-Q.}\
  \bibnamefont {Chen}}, \bibinfo {author} {\bibfnamefont {D.}~\bibnamefont
  {Yarotski}}, \ and\ \bibinfo {author} {\bibfnamefont {Q.}~\bibnamefont
  {Jia}},\ }\href@noop {} {\bibfield  {journal} {\bibinfo  {journal} {Sci.
  Adv.}\ }\textbf {\bibinfo {volume} {2}},\ \bibinfo {pages} {e1600245}
  (\bibinfo {year} {2016})}\BibitemShut {NoStop}%
\bibitem [{\citenamefont {Ohtomo}\ and\ \citenamefont
  {Hwang}(2004)}]{Ohtomo2004}%
  \BibitemOpen
  \bibfield  {author} {\bibinfo {author} {\bibfnamefont {A.}~\bibnamefont
  {Ohtomo}}\ and\ \bibinfo {author} {\bibfnamefont {H.~Y.}\ \bibnamefont
  {Hwang}},\ }\href@noop {} {\bibfield  {journal} {\bibinfo  {journal}
  {Nature}\ }\textbf {\bibinfo {volume} {427}},\ \bibinfo {pages} {423}
  (\bibinfo {year} {2004})}\BibitemShut {NoStop}%
\bibitem [{\citenamefont {Reyren}\ \emph {et~al.}(2007)\citenamefont {Reyren},
  \citenamefont {Thiel}, \citenamefont {Caviglia}, \citenamefont {Kourkoutis},
  \citenamefont {Hammerl}, \citenamefont {Richter}, \citenamefont {Schneider},
  \citenamefont {Kopp}, \citenamefont {R{\"u}etschi}, \citenamefont {Jaccard},
  \citenamefont {Gabay}, \citenamefont {Muller}, \citenamefont {Triscone},\
  and\ \citenamefont {Mannhart}}]{Reyren1196}%
  \BibitemOpen
  \bibfield  {author} {\bibinfo {author} {\bibfnamefont {N.}~\bibnamefont
  {Reyren}}, \bibinfo {author} {\bibfnamefont {S.}~\bibnamefont {Thiel}},
  \bibinfo {author} {\bibfnamefont {A.~D.}\ \bibnamefont {Caviglia}}, \bibinfo
  {author} {\bibfnamefont {L.~F.}\ \bibnamefont {Kourkoutis}}, \bibinfo
  {author} {\bibfnamefont {G.}~\bibnamefont {Hammerl}}, \bibinfo {author}
  {\bibfnamefont {C.}~\bibnamefont {Richter}}, \bibinfo {author} {\bibfnamefont
  {C.~W.}\ \bibnamefont {Schneider}}, \bibinfo {author} {\bibfnamefont
  {T.}~\bibnamefont {Kopp}}, \bibinfo {author} {\bibfnamefont {A.-S.}\
  \bibnamefont {R{\"u}etschi}}, \bibinfo {author} {\bibfnamefont
  {D.}~\bibnamefont {Jaccard}}, \bibinfo {author} {\bibfnamefont
  {M.}~\bibnamefont {Gabay}}, \bibinfo {author} {\bibfnamefont {D.~A.}\
  \bibnamefont {Muller}}, \bibinfo {author} {\bibfnamefont {J.-M.}\
  \bibnamefont {Triscone}}, \ and\ \bibinfo {author} {\bibfnamefont
  {J.}~\bibnamefont {Mannhart}},\ }\href {\doibase 10.1126/science.1146006}
  {\bibfield  {journal} {\bibinfo  {journal} {Science}\ }\textbf {\bibinfo
  {volume} {317}},\ \bibinfo {pages} {1196} (\bibinfo {year}
  {2007})}\BibitemShut {NoStop}%
\bibitem [{\citenamefont {Ben~Shalom}\ \emph {et~al.}(2010)\citenamefont
  {Ben~Shalom}, \citenamefont {Sachs}, \citenamefont {Rakhmilevitch},
  \citenamefont {Palevski},\ and\ \citenamefont {Dagan}}]{Shalom2010}%
  \BibitemOpen
  \bibfield  {author} {\bibinfo {author} {\bibfnamefont {M.}~\bibnamefont
  {Ben~Shalom}}, \bibinfo {author} {\bibfnamefont {M.}~\bibnamefont {Sachs}},
  \bibinfo {author} {\bibfnamefont {D.}~\bibnamefont {Rakhmilevitch}}, \bibinfo
  {author} {\bibfnamefont {A.}~\bibnamefont {Palevski}}, \ and\ \bibinfo
  {author} {\bibfnamefont {Y.}~\bibnamefont {Dagan}},\ }\href {\doibase
  10.1103/PhysRevLett.104.126802} {\bibfield  {journal} {\bibinfo  {journal}
  {Phys. Rev. Lett.}\ }\textbf {\bibinfo {volume} {104}},\ \bibinfo {pages}
  {126802} (\bibinfo {year} {2010})}\BibitemShut {NoStop}%
\bibitem [{\citenamefont {Wang}\ \emph {et~al.}(2012)\citenamefont {Wang},
  \citenamefont {Kalantar-Zadeh}, \citenamefont {Kis}, \citenamefont
  {Coleman},\ and\ \citenamefont {Strano}}]{Wang2012}%
  \BibitemOpen
  \bibfield  {author} {\bibinfo {author} {\bibfnamefont {Q.~H.}\ \bibnamefont
  {Wang}}, \bibinfo {author} {\bibfnamefont {K.}~\bibnamefont
  {Kalantar-Zadeh}}, \bibinfo {author} {\bibfnamefont {A.}~\bibnamefont {Kis}},
  \bibinfo {author} {\bibfnamefont {J.~N.}\ \bibnamefont {Coleman}}, \ and\
  \bibinfo {author} {\bibfnamefont {M.~S.}\ \bibnamefont {Strano}},\
  }\href@noop {} {\bibfield  {journal} {\bibinfo  {journal} {Nat.
  Nanotechnol.}\ }\textbf {\bibinfo {volume} {7}},\ \bibinfo {pages} {699}
  (\bibinfo {year} {2012})}\BibitemShut {NoStop}%
\bibitem [{\citenamefont {Kane}\ and\ \citenamefont {Mele}(2005)}]{KaneMele}%
  \BibitemOpen
  \bibfield  {author} {\bibinfo {author} {\bibfnamefont {C.~L.}\ \bibnamefont
  {Kane}}\ and\ \bibinfo {author} {\bibfnamefont {E.~J.}\ \bibnamefont
  {Mele}},\ }\href {\doibase 10.1103/PhysRevLett.95.226801} {\bibfield
  {journal} {\bibinfo  {journal} {Phys. Rev. Lett.}\ }\textbf {\bibinfo
  {volume} {95}},\ \bibinfo {pages} {226801} (\bibinfo {year}
  {2005})}\BibitemShut {NoStop}%
\bibitem [{\citenamefont {Kobayashi}\ \emph {et~al.}(1998)\citenamefont
  {Kobayashi}, \citenamefont {Kimura}, \citenamefont {Sawada}, \citenamefont
  {Terakura},\ and\ \citenamefont {Tokura}}]{SFMOKobayashi}%
  \BibitemOpen
  \bibfield  {author} {\bibinfo {author} {\bibfnamefont {K.-I.}\ \bibnamefont
  {Kobayashi}}, \bibinfo {author} {\bibfnamefont {T.}~\bibnamefont {Kimura}},
  \bibinfo {author} {\bibfnamefont {H.}~\bibnamefont {Sawada}}, \bibinfo
  {author} {\bibfnamefont {K.}~\bibnamefont {Terakura}}, \ and\ \bibinfo
  {author} {\bibfnamefont {Y.}~\bibnamefont {Tokura}},\ }\href@noop {}
  {\bibfield  {journal} {\bibinfo  {journal} {Nature}\ }\textbf {\bibinfo
  {volume} {395}},\ \bibinfo {pages} {677} (\bibinfo {year}
  {1998})}\BibitemShut {NoStop}%
\bibitem [{\citenamefont {Maignan}\ \emph {et~al.}(1999)\citenamefont
  {Maignan}, \citenamefont {Raveau}, \citenamefont {Martin},\ and\
  \citenamefont {Hervieu}}]{MAIGNAN1999224}%
  \BibitemOpen
  \bibfield  {author} {\bibinfo {author} {\bibfnamefont {A.}~\bibnamefont
  {Maignan}}, \bibinfo {author} {\bibfnamefont {B.}~\bibnamefont {Raveau}},
  \bibinfo {author} {\bibfnamefont {C.}~\bibnamefont {Martin}}, \ and\ \bibinfo
  {author} {\bibfnamefont {M.}~\bibnamefont {Hervieu}},\ }\href@noop {}
  {\bibfield  {journal} {\bibinfo  {journal} {J. Solid State Chem.}\ }\textbf
  {\bibinfo {volume} {144}},\ \bibinfo {pages} {224} (\bibinfo {year}
  {1999})}\BibitemShut {NoStop}%
\bibitem [{\citenamefont {Borges}\ \emph {et~al.}(1999)\citenamefont {Borges},
  \citenamefont {Thomas}, \citenamefont {Cullinan}, \citenamefont {Coey},
  \citenamefont {Suryanarayanan}, \citenamefont {Ben-Dor}, \citenamefont
  {Pinsard-Gaudart},\ and\ \citenamefont {Revcolevschi}}]{AFMO}%
  \BibitemOpen
  \bibfield  {author} {\bibinfo {author} {\bibfnamefont {R.~P.}\ \bibnamefont
  {Borges}}, \bibinfo {author} {\bibfnamefont {R.~M.}\ \bibnamefont {Thomas}},
  \bibinfo {author} {\bibfnamefont {C.}~\bibnamefont {Cullinan}}, \bibinfo
  {author} {\bibfnamefont {J.~M.~D.}\ \bibnamefont {Coey}}, \bibinfo {author}
  {\bibfnamefont {R.}~\bibnamefont {Suryanarayanan}}, \bibinfo {author}
  {\bibfnamefont {L.}~\bibnamefont {Ben-Dor}}, \bibinfo {author} {\bibfnamefont
  {L.}~\bibnamefont {Pinsard-Gaudart}}, \ and\ \bibinfo {author} {\bibfnamefont
  {A.}~\bibnamefont {Revcolevschi}},\ }\href@noop {} {\bibfield  {journal}
  {\bibinfo  {journal} {J. Phys.: Condens. Matter}\ }\textbf {\bibinfo {volume}
  {11}},\ \bibinfo {pages} {L445} (\bibinfo {year} {1999})}\BibitemShut
  {NoStop}%
\bibitem [{\citenamefont {Yuan}\ \emph {et~al.}(2013)\citenamefont {Yuan},
  \citenamefont {Xu},\ and\ \citenamefont {Chen}}]{PFMO}%
  \BibitemOpen
  \bibfield  {author} {\bibinfo {author} {\bibfnamefont {X.}~\bibnamefont
  {Yuan}}, \bibinfo {author} {\bibfnamefont {M.}~\bibnamefont {Xu}}, \ and\
  \bibinfo {author} {\bibfnamefont {Y.}~\bibnamefont {Chen}},\ }\href@noop {}
  {\bibfield  {journal} {\bibinfo  {journal} {Appl. Phys. Lett.}\ }\textbf
  {\bibinfo {volume} {103}},\ \bibinfo {pages} {052411} (\bibinfo {year}
  {2013})}\BibitemShut {NoStop}%
\bibitem [{\citenamefont {Sarma}\ \emph {et~al.}(2000)\citenamefont {Sarma},
  \citenamefont {Mahadevan}, \citenamefont {Saha-Dasgupta}, \citenamefont
  {Ray},\ and\ \citenamefont {Kumar}}]{Sarma2000}%
  \BibitemOpen
  \bibfield  {author} {\bibinfo {author} {\bibfnamefont {D.~D.}\ \bibnamefont
  {Sarma}}, \bibinfo {author} {\bibfnamefont {P.}~\bibnamefont {Mahadevan}},
  \bibinfo {author} {\bibfnamefont {T.}~\bibnamefont {Saha-Dasgupta}}, \bibinfo
  {author} {\bibfnamefont {S.}~\bibnamefont {Ray}}, \ and\ \bibinfo {author}
  {\bibfnamefont {A.}~\bibnamefont {Kumar}},\ }\href {\doibase
  10.1103/PhysRevLett.85.2549} {\bibfield  {journal} {\bibinfo  {journal}
  {Phys. Rev. Lett.}\ }\textbf {\bibinfo {volume} {85}},\ \bibinfo {pages}
  {2549} (\bibinfo {year} {2000})}\BibitemShut {NoStop}%
\bibitem [{\citenamefont {Fang}\ \emph {et~al.}(2001)\citenamefont {Fang},
  \citenamefont {Terakura},\ and\ \citenamefont {Kanamori}}]{Fang2001}%
  \BibitemOpen
  \bibfield  {author} {\bibinfo {author} {\bibfnamefont {Z.}~\bibnamefont
  {Fang}}, \bibinfo {author} {\bibfnamefont {K.}~\bibnamefont {Terakura}}, \
  and\ \bibinfo {author} {\bibfnamefont {J.}~\bibnamefont {Kanamori}},\ }\href
  {\doibase 10.1103/PhysRevB.63.180407} {\bibfield  {journal} {\bibinfo
  {journal} {Phys. Rev. B}\ }\textbf {\bibinfo {volume} {63}},\ \bibinfo
  {pages} {180407} (\bibinfo {year} {2001})}\BibitemShut {NoStop}%
\bibitem [{\citenamefont {Solovyev}(2002)}]{Solo2002}%
  \BibitemOpen
  \bibfield  {author} {\bibinfo {author} {\bibfnamefont {I.~V.}\ \bibnamefont
  {Solovyev}},\ }\href {\doibase 10.1103/PhysRevB.65.144446} {\bibfield
  {journal} {\bibinfo  {journal} {Phys. Rev. B}\ }\textbf {\bibinfo {volume}
  {65}},\ \bibinfo {pages} {144446} (\bibinfo {year} {2002})}\BibitemShut
  {NoStop}%
\bibitem [{\citenamefont {Saitoh}\ \emph {et~al.}(2002)\citenamefont {Saitoh},
  \citenamefont {Nakatake}, \citenamefont {Kakizaki}, \citenamefont {Nakajima},
  \citenamefont {Morimoto}, \citenamefont {Xu}, \citenamefont {Moritomo},
  \citenamefont {Hamada},\ and\ \citenamefont {Aiura}}]{Sait2002}%
  \BibitemOpen
  \bibfield  {author} {\bibinfo {author} {\bibfnamefont {T.}~\bibnamefont
  {Saitoh}}, \bibinfo {author} {\bibfnamefont {M.}~\bibnamefont {Nakatake}},
  \bibinfo {author} {\bibfnamefont {A.}~\bibnamefont {Kakizaki}}, \bibinfo
  {author} {\bibfnamefont {H.}~\bibnamefont {Nakajima}}, \bibinfo {author}
  {\bibfnamefont {O.}~\bibnamefont {Morimoto}}, \bibinfo {author}
  {\bibfnamefont {S.}~\bibnamefont {Xu}}, \bibinfo {author} {\bibfnamefont
  {Y.}~\bibnamefont {Moritomo}}, \bibinfo {author} {\bibfnamefont
  {N.}~\bibnamefont {Hamada}}, \ and\ \bibinfo {author} {\bibfnamefont
  {Y.}~\bibnamefont {Aiura}},\ }\href {\doibase 10.1103/PhysRevB.66.035112}
  {\bibfield  {journal} {\bibinfo  {journal} {Phys. Rev. B}\ }\textbf {\bibinfo
  {volume} {66}},\ \bibinfo {pages} {035112} (\bibinfo {year}
  {2002})}\BibitemShut {NoStop}%
\bibitem [{\citenamefont {Chattopadhyay}\ and\ \citenamefont
  {Millis}(2001)}]{Chat2001}%
  \BibitemOpen
  \bibfield  {author} {\bibinfo {author} {\bibfnamefont {A.}~\bibnamefont
  {Chattopadhyay}}\ and\ \bibinfo {author} {\bibfnamefont {A.~J.}\ \bibnamefont
  {Millis}},\ }\href {\doibase 10.1103/PhysRevB.64.024424} {\bibfield
  {journal} {\bibinfo  {journal} {Phys. Rev. B}\ }\textbf {\bibinfo {volume}
  {64}},\ \bibinfo {pages} {024424} (\bibinfo {year} {2001})}\BibitemShut
  {NoStop}%
\bibitem [{\citenamefont {Jackeli}(2003)}]{Jackeli}%
  \BibitemOpen
  \bibfield  {author} {\bibinfo {author} {\bibfnamefont {G.}~\bibnamefont
  {Jackeli}},\ }\href@noop {} {\bibfield  {journal} {\bibinfo  {journal} {Phys.
  Rev. B}\ }\textbf {\bibinfo {volume} {68}},\ \bibinfo {pages} {092401}
  (\bibinfo {year} {2003})}\BibitemShut {NoStop}%
\bibitem [{\citenamefont {Erten}\ \emph {et~al.}(2011)\citenamefont {Erten},
  \citenamefont {Meetei}, \citenamefont {Mukherjee}, \citenamefont {Randeria},
  \citenamefont {Trivedi},\ and\ \citenamefont {Woodward}}]{Onur2011}%
  \BibitemOpen
  \bibfield  {author} {\bibinfo {author} {\bibfnamefont {O.}~\bibnamefont
  {Erten}}, \bibinfo {author} {\bibfnamefont {O.~N.}\ \bibnamefont {Meetei}},
  \bibinfo {author} {\bibfnamefont {A.}~\bibnamefont {Mukherjee}}, \bibinfo
  {author} {\bibfnamefont {M.}~\bibnamefont {Randeria}}, \bibinfo {author}
  {\bibfnamefont {N.}~\bibnamefont {Trivedi}}, \ and\ \bibinfo {author}
  {\bibfnamefont {P.}~\bibnamefont {Woodward}},\ }\href {\doibase
  10.1103/PhysRevLett.107.257201} {\bibfield  {journal} {\bibinfo  {journal}
  {Phys. Rev. Lett.}\ }\textbf {\bibinfo {volume} {107}},\ \bibinfo {pages}
  {257201} (\bibinfo {year} {2011})}\BibitemShut {NoStop}%
\bibitem [{\citenamefont {Du}\ \emph {et~al.}(2013)\citenamefont {Du},
  \citenamefont {Adur}, \citenamefont {Wang}, \citenamefont {Hauser},
  \citenamefont {Yang},\ and\ \citenamefont {Hammel}}]{substrate}%
  \BibitemOpen
  \bibfield  {author} {\bibinfo {author} {\bibfnamefont {C.}~\bibnamefont
  {Du}}, \bibinfo {author} {\bibfnamefont {R.}~\bibnamefont {Adur}}, \bibinfo
  {author} {\bibfnamefont {H.}~\bibnamefont {Wang}}, \bibinfo {author}
  {\bibfnamefont {A.~J.}\ \bibnamefont {Hauser}}, \bibinfo {author}
  {\bibfnamefont {F.}~\bibnamefont {Yang}}, \ and\ \bibinfo {author}
  {\bibfnamefont {P.~C.}\ \bibnamefont {Hammel}},\ }\href {\doibase
  10.1103/PhysRevLett.110.147204} {\bibfield  {journal} {\bibinfo  {journal}
  {Phys. Rev. Lett.}\ }\textbf {\bibinfo {volume} {110}},\ \bibinfo {pages}
  {147204} (\bibinfo {year} {2013})}\BibitemShut {NoStop}%
\bibitem [{\citenamefont {Hauser}\ \emph {et~al.}(2011)\citenamefont {Hauser},
  \citenamefont {Williams}, \citenamefont {Ricciardo}, \citenamefont {Genc},
  \citenamefont {Dixit}, \citenamefont {Lucy}, \citenamefont {Woodward},
  \citenamefont {Fraser},\ and\ \citenamefont {Yang}}]{Hauser}%
  \BibitemOpen
  \bibfield  {author} {\bibinfo {author} {\bibfnamefont {A.~J.}\ \bibnamefont
  {Hauser}}, \bibinfo {author} {\bibfnamefont {R.~E.~A.}\ \bibnamefont
  {Williams}}, \bibinfo {author} {\bibfnamefont {R.~A.}\ \bibnamefont
  {Ricciardo}}, \bibinfo {author} {\bibfnamefont {A.}~\bibnamefont {Genc}},
  \bibinfo {author} {\bibfnamefont {M.}~\bibnamefont {Dixit}}, \bibinfo
  {author} {\bibfnamefont {J.~M.}\ \bibnamefont {Lucy}}, \bibinfo {author}
  {\bibfnamefont {P.~M.}\ \bibnamefont {Woodward}}, \bibinfo {author}
  {\bibfnamefont {H.~L.}\ \bibnamefont {Fraser}}, \ and\ \bibinfo {author}
  {\bibfnamefont {F.}~\bibnamefont {Yang}},\ }\href@noop {} {\bibfield
  {journal} {\bibinfo  {journal} {Phys. Rev. B}\ }\textbf {\bibinfo {volume}
  {83}},\ \bibinfo {pages} {014407} (\bibinfo {year} {2011})}\BibitemShut
  {NoStop}%
\bibitem [{\citenamefont {Chen}\ \emph {et~al.}(2010)\citenamefont {Chen},
  \citenamefont {Pereira},\ and\ \citenamefont {Balents}}]{Balents}%
  \BibitemOpen
  \bibfield  {author} {\bibinfo {author} {\bibfnamefont {G.}~\bibnamefont
  {Chen}}, \bibinfo {author} {\bibfnamefont {R.}~\bibnamefont {Pereira}}, \
  and\ \bibinfo {author} {\bibfnamefont {L.}~\bibnamefont {Balents}},\
  }\href@noop {} {\bibfield  {journal} {\bibinfo  {journal} {Phys. Rev. B}\
  }\textbf {\bibinfo {volume} {82}},\ \bibinfo {pages} {174440} (\bibinfo
  {year} {2010})}\BibitemShut {NoStop}%
\bibitem [{\citenamefont {Natori}\ \emph {et~al.}(2016)\citenamefont {Natori},
  \citenamefont {Andrade}, \citenamefont {Miranda},\ and\ \citenamefont
  {Pereira}}]{Natori2016}%
  \BibitemOpen
  \bibfield  {author} {\bibinfo {author} {\bibfnamefont {W.~M.~H.}\
  \bibnamefont {Natori}}, \bibinfo {author} {\bibfnamefont {E.~C.}\
  \bibnamefont {Andrade}}, \bibinfo {author} {\bibfnamefont {E.}~\bibnamefont
  {Miranda}}, \ and\ \bibinfo {author} {\bibfnamefont {R.~G.}\ \bibnamefont
  {Pereira}},\ }\href {\doibase 10.1103/PhysRevLett.117.017204} {\bibfield
  {journal} {\bibinfo  {journal} {Phys. Rev. Lett.}\ }\textbf {\bibinfo
  {volume} {117}},\ \bibinfo {pages} {017204} (\bibinfo {year}
  {2016})}\BibitemShut {NoStop}%
\bibitem [{\citenamefont {Romh\'anyi}\ \emph {et~al.}(2017)\citenamefont
  {Romh\'anyi}, \citenamefont {Balents},\ and\ \citenamefont
  {Jackeli}}]{Romh2016}%
  \BibitemOpen
  \bibfield  {author} {\bibinfo {author} {\bibfnamefont {J.}~\bibnamefont
  {Romh\'anyi}}, \bibinfo {author} {\bibfnamefont {L.}~\bibnamefont {Balents}},
  \ and\ \bibinfo {author} {\bibfnamefont {G.}~\bibnamefont {Jackeli}},\ }\href
  {\doibase 10.1103/PhysRevLett.118.217202} {\bibfield  {journal} {\bibinfo
  {journal} {Phys. Rev. Lett.}\ }\textbf {\bibinfo {volume} {118}},\ \bibinfo
  {pages} {217202} (\bibinfo {year} {2017})}\BibitemShut {NoStop}%
\bibitem [{\citenamefont {Chen}\ and\ \citenamefont
  {Balents}(2008)}]{Chen2008}%
  \BibitemOpen
  \bibfield  {author} {\bibinfo {author} {\bibfnamefont {G.}~\bibnamefont
  {Chen}}\ and\ \bibinfo {author} {\bibfnamefont {L.}~\bibnamefont {Balents}},\
  }\href {\doibase 10.1103/PhysRevB.78.094403} {\bibfield  {journal} {\bibinfo
  {journal} {Phys. Rev. B}\ }\textbf {\bibinfo {volume} {78}},\ \bibinfo
  {pages} {094403} (\bibinfo {year} {2008})}\BibitemShut {NoStop}%
\bibitem [{\citenamefont {Jackeli}\ and\ \citenamefont
  {Khaliullin}(2009{\natexlab{a}})}]{Jackeli2009}%
  \BibitemOpen
  \bibfield  {author} {\bibinfo {author} {\bibfnamefont {G.}~\bibnamefont
  {Jackeli}}\ and\ \bibinfo {author} {\bibfnamefont {G.}~\bibnamefont
  {Khaliullin}},\ }\href {\doibase 10.1103/PhysRevLett.102.017205} {\bibfield
  {journal} {\bibinfo  {journal} {Phys. Rev. Lett.}\ }\textbf {\bibinfo
  {volume} {102}},\ \bibinfo {pages} {017205} (\bibinfo {year}
  {2009}{\natexlab{a}})}\BibitemShut {NoStop}%
\bibitem [{\citenamefont {Cook}\ and\ \citenamefont
  {Paramekanti}(2014)}]{Cook}%
  \BibitemOpen
  \bibfield  {author} {\bibinfo {author} {\bibfnamefont {A.~M.}\ \bibnamefont
  {Cook}}\ and\ \bibinfo {author} {\bibfnamefont {A.}~\bibnamefont
  {Paramekanti}},\ }\href@noop {} {\bibfield  {journal} {\bibinfo  {journal}
  {Phys. Rev. Lett.}\ }\textbf {\bibinfo {volume} {113}},\ \bibinfo {pages}
  {077203} (\bibinfo {year} {2014})}\BibitemShut {NoStop}%
\bibitem [{\citenamefont {Cook}(2016)}]{AMCook2016}%
  \BibitemOpen
  \bibfield  {author} {\bibinfo {author} {\bibfnamefont {A.~M.}\ \bibnamefont
  {Cook}},\ }\href {\doibase 10.1103/PhysRevB.94.205135} {\bibfield  {journal}
  {\bibinfo  {journal} {Phys. Rev. B}\ }\textbf {\bibinfo {volume} {94}},\
  \bibinfo {pages} {205135} (\bibinfo {year} {2016})}\BibitemShut {NoStop}%
\bibitem [{\citenamefont {Baidya}\ \emph {et~al.}(2015)\citenamefont {Baidya},
  \citenamefont {Waghmare}, \citenamefont {Paramekanti},\ and\ \citenamefont
  {Saha-Dasgupta}}]{BFRO}%
  \BibitemOpen
  \bibfield  {author} {\bibinfo {author} {\bibfnamefont {S.}~\bibnamefont
  {Baidya}}, \bibinfo {author} {\bibfnamefont {U.~V.}\ \bibnamefont
  {Waghmare}}, \bibinfo {author} {\bibfnamefont {A.}~\bibnamefont
  {Paramekanti}}, \ and\ \bibinfo {author} {\bibfnamefont {T.}~\bibnamefont
  {Saha-Dasgupta}},\ }\href@noop {} {\bibfield  {journal} {\bibinfo  {journal}
  {Phys. Rev. B}\ }\textbf {\bibinfo {volume} {92}},\ \bibinfo {pages} {161106}
  (\bibinfo {year} {2015})}\BibitemShut {NoStop}%
\bibitem [{\citenamefont {Jackeli}\ and\ \citenamefont
  {Khaliullin}(2009{\natexlab{b}})}]{SVO}%
  \BibitemOpen
  \bibfield  {author} {\bibinfo {author} {\bibfnamefont {G.}~\bibnamefont
  {Jackeli}}\ and\ \bibinfo {author} {\bibfnamefont {G.}~\bibnamefont
  {Khaliullin}},\ }\href@noop {} {\bibfield  {journal} {\bibinfo  {journal}
  {Phys. Rev. Lett.}\ }\textbf {\bibinfo {volume} {103}},\ \bibinfo {pages}
  {067205} (\bibinfo {year} {2009}{\natexlab{b}})}\BibitemShut {NoStop}%
\bibitem [{\citenamefont {Abragam}\ and\ \citenamefont
  {Bleaney}(1970)}]{Abr70}%
  \BibitemOpen
  \bibfield  {author} {\bibinfo {author} {\bibfnamefont {A.}~\bibnamefont
  {Abragam}}\ and\ \bibinfo {author} {\bibfnamefont {B.}~\bibnamefont
  {Bleaney}},\ }\href@noop {} {\emph {\bibinfo {title} {Electron Paramagnetic
  Resonance of Transition Ions}}}\ (\bibinfo  {publisher} {Clarendon Press,
  Oxford},\ \bibinfo {year} {1970})\BibitemShut {NoStop}%
\bibitem [{\citenamefont {Yamada}\ \emph {et~al.}(2018)\citenamefont {Yamada},
  \citenamefont {Oshikawa},\ and\ \citenamefont {Jackeli}}]{Yamada2018}%
  \BibitemOpen
  \bibfield  {author} {\bibinfo {author} {\bibfnamefont {M.~G.}\ \bibnamefont
  {Yamada}}, \bibinfo {author} {\bibfnamefont {M.}~\bibnamefont {Oshikawa}}, \
  and\ \bibinfo {author} {\bibfnamefont {G.}~\bibnamefont {Jackeli}},\ }\href
  {\doibase 10.1103/PhysRevLett.121.097201} {\bibfield  {journal} {\bibinfo
  {journal} {Phys. Rev. Lett.}\ }\textbf {\bibinfo {volume} {121}},\ \bibinfo
  {pages} {097201} (\bibinfo {year} {2018})}\BibitemShut {NoStop}%
\bibitem [{g()}]{g}%
  \BibitemOpen
  \href@noop {} {}\bibinfo {note} {All matrix elements of the total magnetic
  moment $\vec{M}=2\vec{s}-\vec{l}$ in the pure $j=l+s=3/2$ manifold are
  strictly zero and the corresponding $g$-factor vanishes ($g=0$). However, the
  covalence mixing between the $d$-orbitals and the oxygen $p$-orbitals gives
  rise to a finite but small $g$-factor ($g\ll 1$)~\cite{Abr70}.}\BibitemShut
  {Stop}%
\bibitem [{\citenamefont {Wang}\ \emph {et~al.}(2014)\citenamefont {Wang},
  \citenamefont {Zhong}, \citenamefont {Hao}, \citenamefont {Gerhold},
  \citenamefont {Stöger}, \citenamefont {Schmid}, \citenamefont
  {Sánchez-Barriga}, \citenamefont {Varykhalov}, \citenamefont {Franchini},
  \citenamefont {Held},\ and\ \citenamefont {Diebold}}]{Wang}%
  \BibitemOpen
  \bibfield  {author} {\bibinfo {author} {\bibfnamefont {Z.}~\bibnamefont
  {Wang}}, \bibinfo {author} {\bibfnamefont {Z.}~\bibnamefont {Zhong}},
  \bibinfo {author} {\bibfnamefont {X.}~\bibnamefont {Hao}}, \bibinfo {author}
  {\bibfnamefont {S.}~\bibnamefont {Gerhold}}, \bibinfo {author} {\bibfnamefont
  {B.}~\bibnamefont {Stöger}}, \bibinfo {author} {\bibfnamefont
  {M.}~\bibnamefont {Schmid}}, \bibinfo {author} {\bibfnamefont
  {J.}~\bibnamefont {Sánchez-Barriga}}, \bibinfo {author} {\bibfnamefont
  {A.}~\bibnamefont {Varykhalov}}, \bibinfo {author} {\bibfnamefont
  {C.}~\bibnamefont {Franchini}}, \bibinfo {author} {\bibfnamefont
  {K.}~\bibnamefont {Held}}, \ and\ \bibinfo {author} {\bibfnamefont
  {U.}~\bibnamefont {Diebold}},\ }\href@noop {} {\bibfield  {journal} {\bibinfo
   {journal} {Proc. Natl. Acad. Sci.}\ }\textbf {\bibinfo {volume} {111}},\
  \bibinfo {pages} {3933} (\bibinfo {year} {2014})}\BibitemShut {NoStop}%
\bibitem [{\citenamefont {Garnica}\ \emph {et~al.}(2013)\citenamefont
  {Garnica}, \citenamefont {Stradi}, \citenamefont {Barja}, \citenamefont
  {Calleja}, \citenamefont {D\'iaz}, \citenamefont {Alcam\'i}, \citenamefont
  {Mart\'in}, \citenamefont {V\'azquez~de Parga}, \citenamefont {Mart\'in},\
  and\ \citenamefont {Miranda}}]{Garnica}%
  \BibitemOpen
  \bibfield  {author} {\bibinfo {author} {\bibfnamefont {M.}~\bibnamefont
  {Garnica}}, \bibinfo {author} {\bibfnamefont {D.}~\bibnamefont {Stradi}},
  \bibinfo {author} {\bibfnamefont {S.}~\bibnamefont {Barja}}, \bibinfo
  {author} {\bibfnamefont {F.}~\bibnamefont {Calleja}}, \bibinfo {author}
  {\bibfnamefont {C.}~\bibnamefont {D\'iaz}}, \bibinfo {author} {\bibfnamefont
  {M.}~\bibnamefont {Alcam\'i}}, \bibinfo {author} {\bibfnamefont
  {N.}~\bibnamefont {Mart\'in}}, \bibinfo {author} {\bibfnamefont {A.~L.}\
  \bibnamefont {V\'azquez~de Parga}}, \bibinfo {author} {\bibfnamefont
  {F.}~\bibnamefont {Mart\'in}}, \ and\ \bibinfo {author} {\bibfnamefont
  {R.}~\bibnamefont {Miranda}},\ }\href@noop {} {\bibfield  {journal} {\bibinfo
   {journal} {Nat. Phys.}\ }\textbf {\bibinfo {volume} {9}},\ \bibinfo {pages}
  {368} (\bibinfo {year} {2013})}\BibitemShut {NoStop}%
\bibitem [{\citenamefont {Yamada}\ \emph {et~al.}(2016)\citenamefont {Yamada},
  \citenamefont {Soejima}, \citenamefont {Tsuji}, \citenamefont {Hirai},
  \citenamefont {Dinc\u{a}},\ and\ \citenamefont {Aoki}}]{Yamada}%
  \BibitemOpen
  \bibfield  {author} {\bibinfo {author} {\bibfnamefont {M.~G.}\ \bibnamefont
  {Yamada}}, \bibinfo {author} {\bibfnamefont {T.}~\bibnamefont {Soejima}},
  \bibinfo {author} {\bibfnamefont {N.}~\bibnamefont {Tsuji}}, \bibinfo
  {author} {\bibfnamefont {D.}~\bibnamefont {Hirai}}, \bibinfo {author}
  {\bibfnamefont {M.}~\bibnamefont {Dinc\u{a}}}, \ and\ \bibinfo {author}
  {\bibfnamefont {H.}~\bibnamefont {Aoki}},\ }\href@noop {} {\bibfield
  {journal} {\bibinfo  {journal} {Phys. Rev. B}\ }\textbf {\bibinfo {volume}
  {94}},\ \bibinfo {pages} {081102} (\bibinfo {year} {2016})}\BibitemShut
  {NoStop}%
\bibitem [{\citenamefont {Lieb}(1989)}]{Lieb}%
  \BibitemOpen
  \bibfield  {author} {\bibinfo {author} {\bibfnamefont {E.~H.}\ \bibnamefont
  {Lieb}},\ }\href@noop {} {\bibfield  {journal} {\bibinfo  {journal} {Phys.
  Rev. Lett.}\ }\textbf {\bibinfo {volume} {62}},\ \bibinfo {pages} {1201}
  (\bibinfo {year} {1989})}\BibitemShut {NoStop}%
\bibitem [{\citenamefont {Mielke}(1991)}]{Mielke}%
  \BibitemOpen
  \bibfield  {author} {\bibinfo {author} {\bibfnamefont {A.}~\bibnamefont
  {Mielke}},\ }\href@noop {} {\bibfield  {journal} {\bibinfo  {journal} {J.
  Phys. A: Math. Gen.}\ }\textbf {\bibinfo {volume} {24}},\ \bibinfo {pages}
  {L73} (\bibinfo {year} {1991})}\BibitemShut {NoStop}%
\bibitem [{\citenamefont {Tasaki}(1992)}]{Tasaki}%
  \BibitemOpen
  \bibfield  {author} {\bibinfo {author} {\bibfnamefont {H.}~\bibnamefont
  {Tasaki}},\ }\href@noop {} {\bibfield  {journal} {\bibinfo  {journal} {Phys.
  Rev. Lett.}\ }\textbf {\bibinfo {volume} {69}},\ \bibinfo {pages} {1608}
  (\bibinfo {year} {1992})}\BibitemShut {NoStop}%
\bibitem [{\citenamefont {Wigner}(1934)}]{Wigner}%
  \BibitemOpen
  \bibfield  {author} {\bibinfo {author} {\bibfnamefont {E.}~\bibnamefont
  {Wigner}},\ }\href@noop {} {\bibfield  {journal} {\bibinfo  {journal} {Phys.
  Rev.}\ }\textbf {\bibinfo {volume} {46}},\ \bibinfo {pages} {1002} (\bibinfo
  {year} {1934})}\BibitemShut {NoStop}%
\bibitem [{\citenamefont {S\'anchez}\ \emph {et~al.}(2004)\citenamefont
  {S\'anchez}, \citenamefont {Alonso}, \citenamefont {Garc\'ia-Hern\'andez},
  \citenamefont {Mart\'inez-Lope}, \citenamefont {Casais}, \citenamefont
  {Mart\'inez},\ and\ \citenamefont {Fern\'andez-D\'iaz}}]{Sanchez}%
  \BibitemOpen
  \bibfield  {author} {\bibinfo {author} {\bibfnamefont {D.}~\bibnamefont
  {S\'anchez}}, \bibinfo {author} {\bibfnamefont {J.~A.}\ \bibnamefont
  {Alonso}}, \bibinfo {author} {\bibfnamefont {M.}~\bibnamefont
  {Garc\'ia-Hern\'andez}}, \bibinfo {author} {\bibfnamefont {M.~J.}\
  \bibnamefont {Mart\'inez-Lope}}, \bibinfo {author} {\bibfnamefont {M.~T.}\
  \bibnamefont {Casais}}, \bibinfo {author} {\bibfnamefont {J.~L.}\
  \bibnamefont {Mart\'inez}}, \ and\ \bibinfo {author} {\bibfnamefont {M.~T.}\
  \bibnamefont {Fern\'andez-D\'iaz}},\ }\href@noop {} {\bibfield  {journal}
  {\bibinfo  {journal} {J. Magn. Magn. Mater.}\ }\textbf {\bibinfo {volume}
  {272--276, Part 3}},\ \bibinfo {pages} {1732} (\bibinfo {year}
  {2004})}\BibitemShut {NoStop}%
\bibitem [{\citenamefont {Chitra}\ \emph {et~al.}(2001)\citenamefont {Chitra},
  \citenamefont {Giamarchi},\ and\ \citenamefont {Le~Doussal}}]{Chitra2001}%
  \BibitemOpen
  \bibfield  {author} {\bibinfo {author} {\bibfnamefont {R.}~\bibnamefont
  {Chitra}}, \bibinfo {author} {\bibfnamefont {T.}~\bibnamefont {Giamarchi}}, \
  and\ \bibinfo {author} {\bibfnamefont {P.}~\bibnamefont {Le~Doussal}},\
  }\href {\doibase 10.1103/PhysRevB.65.035312} {\bibfield  {journal} {\bibinfo
  {journal} {Phys. Rev. B}\ }\textbf {\bibinfo {volume} {65}},\ \bibinfo
  {pages} {035312} (\bibinfo {year} {2001})}\BibitemShut {NoStop}%
\end{thebibliography}%

\end{document}